\pdfoutput=1 
\documentclass{JINST}
\usepackage{cite}
\usepackage{soul}
\usepackage{epstopdf}
\usepackage{multirow}
\usepackage{float}
\title{Characterization of PARIS  LaBr$_3$(Ce)-NaI(Tl) phoswich detectors upto $E_\gamma$ $\sim$~22~MeV}
\author{C.~Ghosh$^a$, V.~Nanal$^a$\thanks{Corresponding author.}, R.G.~Pillay$^a$, Anoop~K.V.$^b$, N.~Dokania$^c$, Sanjoy~Pal$^b$, M.S.~Pose$^a$, G.~Mishra$^d$, P.C.~Rout$^d$, Suresh~Kumar$^d$, Deepak~Pandit$^e$, Debasish~Mondal$^e$, Surajit~Pal$^e$, S.R.~Banerjee$^e$, Pawe{\l}~J.~Napiorkowski$^f$, Oliver~Dorvaux$^g$, S. Kihel$^g$, C.~Mathieu$^g$, A.~Maj$^h$\\
\llap{$^a$}Department of Nuclear and Atomic Physics, Tata Institute of Fundamental Research, Homi Bhabha Road, Mumbai 400 005, India.\\
\llap{$^b$}Pelletron Linac Facility, Tata Institute of Fundamental Research, Homi Bhabha Road, Mumbai 400 005, India.\\
\llap{$^c$}India-based Neutrino Observatory, Tata Institute of Fundamental Research, Homi Bhabha Road, Mumbai 400 005, India.\\
\llap{$^d$}Nuclear Physics Division, Bhabha Atomic Research Centre, Trombay, Mumbai 400 085, India.\\
\llap{$^e$}Variable Energy Cyclotron Centre, 1/AF-Bidhannagar, Kolkata 700 064, India.\\
\llap{$^f$}Heavy Ion Laboratory, University of Warsaw, Pasteura 5a, 02-093 Warsaw, Poland.\\
\llap{$^g$}Institut Pluridisciplinaire Hubert Curien, UNISTRA, CNRS, rue du Loess, 67200 Strasbourg, France.\\
\llap{$^h$}The Niewodnicza{\'{n}}ski Institute of Nuclear Physics, Polish Academy of Sciencies, ul. Radzikowskiego 152, 31-342 Krak{\'{o}}w, Poland.\\
E-mail: \email{nanal@tifr.res.in}}

\abstract{In order to understand the performance of the PARIS (Photon Array for the studies with Radioactive Ion and Stable beams) detector, detailed characterization of two individual phoswich ($\rm LaBr_3(Ce)$-NaI(Tl)) elements has been carried out. The detector response is investigated over a wide range of $E_{\gamma}$~=~0.6~to~22.6~MeV using radioactive sources and employing $^{11}B(p,\gamma)$ reaction at $E_p$ = 163 keV and $E_p$ = 7.2 MeV. The linearity of energy response of the $\rm LaBr_3(Ce)$ detector is tested upto 22.6~MeV using three different voltage dividers.  The data acquisition system using CAEN digitizers is set up and optimized to get the best energy and time resolution.  The energy resolution of $\sim$~2.1\% at $E_\gamma$ = 22.6~MeV is measured for the configuration giving best linearity upto high energy. Time resolution of the phoswich detector is measured with a $\rm ^{60}Co$ source after implementing CFD algorithm for the digitized pulses and is found to be excellent (FWHM~$\sim$~315~ps). In order to study the effect of count rate on detectors, the centroid position and width of the $E_{\gamma}$ = 835~keV peak were measured upto 220~kHz count rate. The measured efficiency data with radioactive sources are in good agreement with GEANT4 based simulations. The total energy spectrum after the add-back of energy signals in phoswich components is also presented.}


\begin{document}
\section{Introduction}\label{}
Scintillation detectors are widely used in Nuclear and Particle Physics research due to their good time resolution, high efficiency and moderate energy resolution. In recent years, rare-earth halide based scintillators ($\rm LaBr_3(Ce)$, $\rm LaCl_3(Ce)$, $\rm CeBr_3$) have found attractive applications \cite{labr,labr1,labr2,labr3,labr5,lacl,cebr,cebr1} due to their excellent energy resolution ($\le 4\%$ at 662~keV), sub-nanosecond time resolution and relatively higher intrinsic efficiency as compared to the conventional scintillators such as NaI(Tl) and BaF$_2$. Amongst these novel scintillators, the $\rm LaBr_3(Ce)$ detector has the best energy resolution \cite{saintgobain} and offers a possible alternative for high resolution gamma ray spectroscopy, particularly for experiments with exotic beams and low-multiplicity events. 

The PARIS (Photon Array for the studies with Radioactive Ion and Stable beams) detector, a high efficiency gamma-calorimeter comprising $\rm LaBr_3(Ce)$-NaI(Tl) phoswich elements (manufactured by Saint-Gobain Crystals), is being developed for the measurement of high energy gamma rays~\cite{paris,parisweb}. The array consists of multiple clusters, each in turn being made of 9 (3$\times$3) phoswich elements. The individual phoswich detector element consists of a front $\rm LaBr_3(Ce)$ crystal ($2^{\prime\prime}\times2^{\prime\prime}\times2^{\prime\prime}$) optically coupled to  a $\rm NaI (Tl)$ crystal ($2^{\prime\prime}\times2^{\prime\prime}\times6^{\prime\prime}$) at the back. Both these crystals are encapsulated together with a single optical window at the rear end, which is coupled to a 8~stage, 46~mm diameter Hamamatsu photomultiplier tube (PMT) R7723-100~\cite{paris1,hamamatsu,hamamatsu1}. This PMT with the maximum sensitivity at $\lambda$ $\sim$~420 nm and a fast pulse rise time $\sim$~1.7~ns is well suited for both  $\rm LaBr_3(Ce)$ ($\lambda_{max}\sim$~380~nm) and NaI(Tl) ($\lambda_{max}\sim$~415~nm). The signals from the $\rm LaBr_3(Ce)$ and NaI(Tl) can be easily resolved due to their widely different decay times, namely, 16~ns and 250~ns, respectively. Ziebli{\'{n}}ski {\it et. al.}~\cite{paris1} have demonstrated the feasibility of separation of individual components with both analog and digital signal processing upto $E_\gamma$ = 6.13 MeV. They have also shown that good energy resolution can be obtained by adding the $\rm LaBr_3(Ce)$ and NaI(Tl) signals.

It should be mentioned that most of the measurements for characterization of $\rm LaBr_3(Ce)$ detectors of different sizes are done with various radioactive sources and hence are restricted to low energy gamma rays. A few high energy measurements, by studying gamma rays with ($p,\gamma$) reactions, have been reported ~\cite{labr1,labr3,labr5}. It is well known that the intense light yield of $\rm LaBr_3(Ce)$ crystal (63~photon/keV$\gamma$~\cite{saintgobain}) may saturate the response of PMT and consequently leads to the non-linearity in energy calibration of the detector. To achieve better linearity over a broad energy range, an active voltage divider has been employed~\cite{labr1,labr5}. Ciema{\l}a et al.~\cite{labr3} have studied the response of a cylindrical $\rm LaBr_3(Ce)$ (2$^{\prime\prime}$ length, 2$^{\prime\prime}$ dia) which is similar to the $\rm LaBr_3(Ce)$ crystal in the phoswich detector element. They have reported energy resolution of 0.7\% and intrinsic photo-peak efficiency of about 0.65\% at 17.6~MeV. They also observed $\sim$~10\% non-linearity at this energy.
  It should be pointed out that for the PARIS detector, the response is further complicated due to the composite nature of the detector and it is important to understand how the performance of the $\rm LaBr_3(Ce)$ is modified by coupling to NaI(Tl). 
 Preliminary high energy gamma ray spectra ($ E_{\gamma}$~=~22.6~MeV) of  the phoswich detector using analog signal processing in $^{11}B(p, \gamma$) reaction were reported in Ref.~\cite{paris1}, which also showed a significant non-linearity. 
Since the readout is common to both $\rm LaBr_3(Ce)$ and NaI(Tl), the optimization for linearity in $\rm LaBr_3(Ce)$ can significantly affect the NaI(Tl) output.  In order to address these issues, a detailed characterization of two phoswich elements have been performed over a wide energy range (0.6 -- 22.6~MeV). 
 Given the composite nature of the pulse, digital processing is best suited for the phoswich detector. In the present study,  the acquisition system using CAEN digitizers (12~bit~250 MS/s, 10~bit~1~GS/s)~\cite{caen} is optimized to get the best energy and time resolution. The energy resolution is extracted by fitting the observed lineshape to the simulated energy spectrum generated using GEANT4~\cite{geant4}. This paper reports the study of the linearity, detector gain stability for high count rates, energy resolution, efficiency and time resolution for the phoswich detector. The total energy spectrum after the add-back of energy signals in phoswich components is presented for high energy gamma rays in $^{11}B(p, \gamma$) reaction. 

\section{Experimental Details}
\begin{figure}[h]
\centering
\includegraphics[scale=0.13]{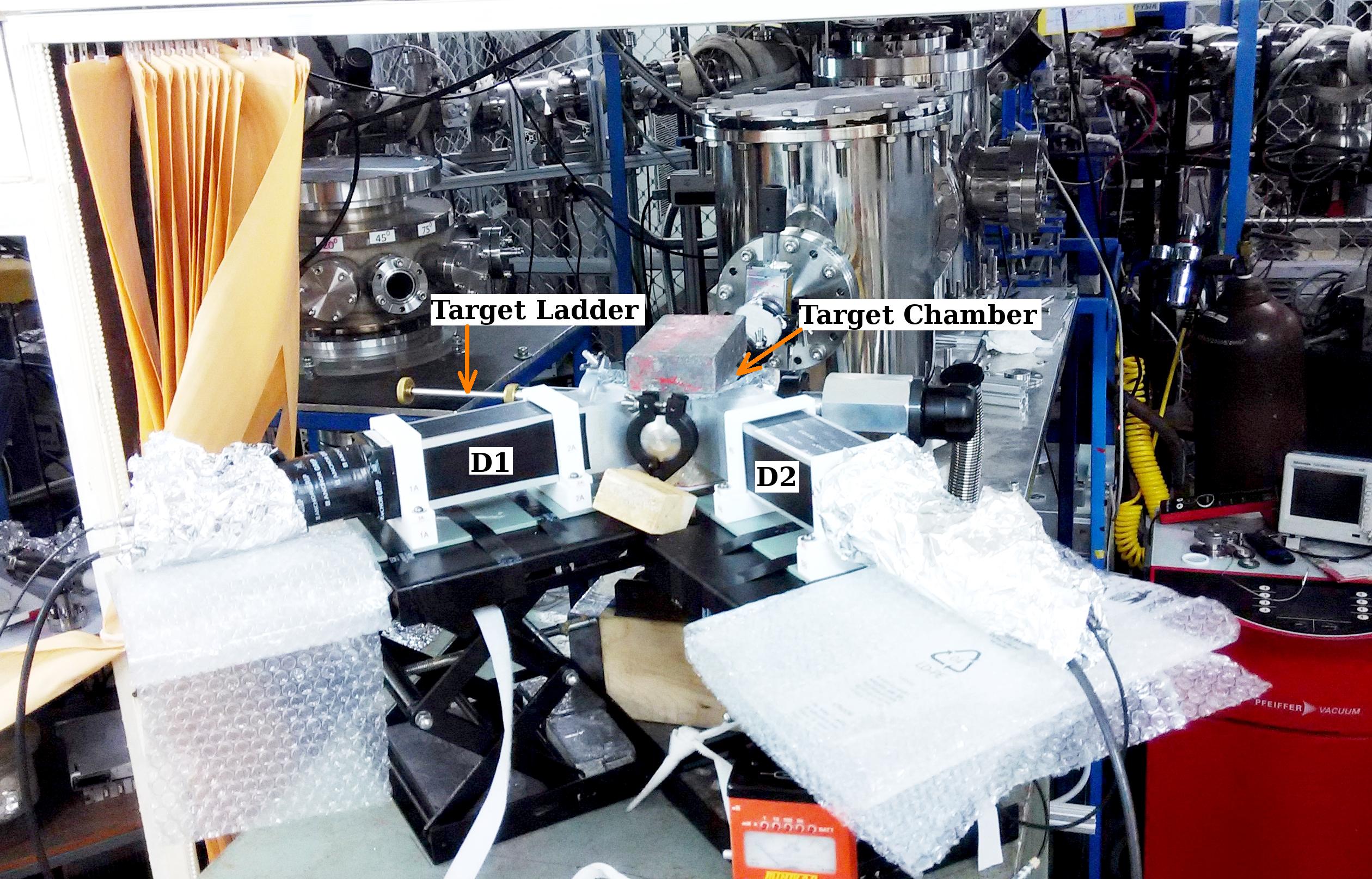}
\caption{\label{setup}(Color online) The experimental set-up used for radiative proton capture reactions.}
\end{figure}

Two phoswich detectors D1 and D2 were coupled to PMT1 and PMT2, respectively, with optical grease.  The ORTEC model 556 unit (with regulation $\leqslant$~0.0025\%, temperature instability~<~$\pm$~50 ppm/$^o$C, long term drift <~0.03\%/24-hour) was used for high voltage bias. Measurements were carried out with  various radioactive sources and the radiative proton capture reaction $^{11}B(p,\gamma)$.
Details of gamma ray energies are given in Table~\ref{source} and Table~\ref{reaction}. The high energy gamma rays were studied  in two separate experiments - first using the 163~keV proton beam from the ECR source~\cite{ecris} and  second using the 7.2~MeV proton beam from the Pelletron Linac Facility, Mumbai~\cite{PLF}. 
Figure~\ref{setup} shows a picture of the experimental set-up used in the radiative capture reactions. A thin walled, compact  Aluminum chamber was used and the detectors were placed at $+45^o$ and $-45^o$ with respect to the beam direction at about 6~cm from the target. The natural Boron target ($\sim1$~mg/cm$^2$) was made by electro-deposition on a 127~$\mu$m thick Tantalum backing. Due to the thick backing the total incident proton charge on the target could not be measured and hence absolute efficiency could not be deduced for high energy gamma rays. Typical current used in the reaction at $E_p$~=~163 keV was $\sim$~2~$\mu$A while that at $E_p$~=~7.2 MeV was about $\sim$10~nA.  A thin Lead absorber ($\sim$~3~mm) was kept in front of both the detectors to absorb X-rays and low energy gamma rays. Adequate shielding was provided to reduce the neutron background in the experiment with $E_p$~=~7.2~MeV. Measurements were carried out over a extended period of time and the detector performance was periodically monitored. 
\begin{table}[H]
\centering
\caption{Radioactive sources used for measurements together with $E_\gamma$~\cite{nndc} and source strengths $N_s$ (as on 1$^{st}$ Feb. 2015).}
\label{source}
\begin{tabular}{ ccc }
\hline
Source & $E_{\gamma}$ (MeV)& $N_s$ (Bq)\\ \hline
$\rm^{22}Na$  & 0.511, 1.2748 & --\\ 
$\rm^{137}Cs$  & 0.6617 & 13738 (427) \\ 
$\rm^{54}Mn$  & 0.8348 & 8260 (298)\\ 
$\rm^{60}Co$ & 1.1732, 1.3325 & 9846 (276)\\ 
$\rm^{241}Am-^9Be$  &4.439 & 9625 (147)\\ 
$\rm^{239}Pu-^{13}C$  &6.130 & --\\  \hline
\end{tabular}
\end{table}
\begin{table}[H]
\centering
\caption{Radiative capture reactions and emitted $E_\gamma$~\cite{pgamma}}
\label{reaction}
\begin{tabular}{  ccc }
\hline
Reaction & $E_{p}$ (MeV) &$E_{\gamma}$ (MeV) \\ \hline
$^{11}B(p,\gamma)$ & 0.163 & 4.439, 11.680\\
\\
$^{11}B(p,\gamma)$ & 7.2  & 4.439, 5.020, \\
& & 18.118, 22.557\\ \hline
\end{tabular}
\end{table}
 As mentioned earlier, for composite pulse from the phoswich detector, the digital signal processing is an ideal choice. However, both analog and digital data acquisition systems were used for part of the measurements ($^{11}B(p, \gamma$) at $E_p$~=~163~keV) for comparison. The analog electronics set-up comprised two charge sensitive ADCs (Silena 4418/Q) with a short gate of 300~ns  (for $\rm LaBr_3(Ce)$ signal) and a long gate of 900~ns (for both $\rm LaBr_3(Ce)$ and NaI(Tl) signals) and data was acquired with LAMPS~\cite{lamps}, a CAMAC-based acquisition and analysis software, similar to preliminary work reported in Ref.~\cite{paris1}.
 
\begin{figure}[h]
\centering
\includegraphics[scale=0.4]{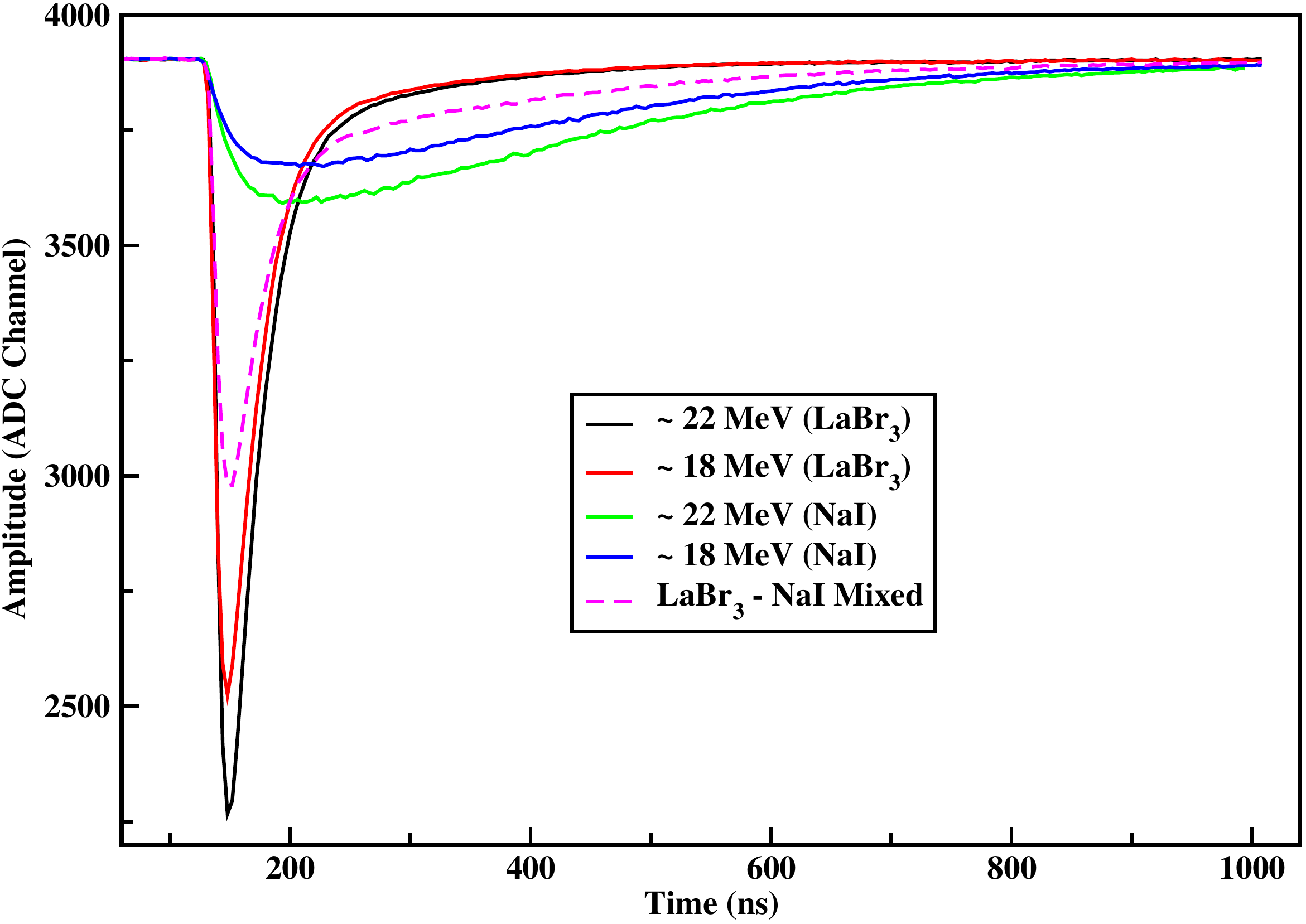}
\caption{\label{pulse}(Color online) High energy gamma ray pulses in detector D1 recorded with V1720e (250~MS/s) digitizer using wavedump software. The mixed signal shown corresponds to an arbitrary  $E_\gamma$ between 18 and 22~MeV. The baseline is set at an arbitrary value around 3900.}
\end{figure}

Two CAEN make VME based digitizers V1720e (12~bit~-~250~MS/s,~2~Volt dynamic range) and V1751 (10~bit~-~1~GS/s,~1~Volt dynamic range)  were used to digitize and process the detector pulse~\cite{caen}. The WAVEDUMP software permits the recording of entire pulse after digitization with settable pre-trigger length.
Typical pulses recorded with V1720e for high energy gamma rays corresponding to different interaction points inside the detector  are shown in Fig.~\ref{pulse}.  The clear separation of events corresponding to full energy deposition in $\rm LaBr_3(Ce)$ or NaI(Tl) as well as partial energy in either of them, is evident. An online algorithm implementing constant fraction discrimination has been developed for timing measurements from the digitized pulse. 

\begin{figure}[ht]
\centering
\includegraphics[scale=0.465]{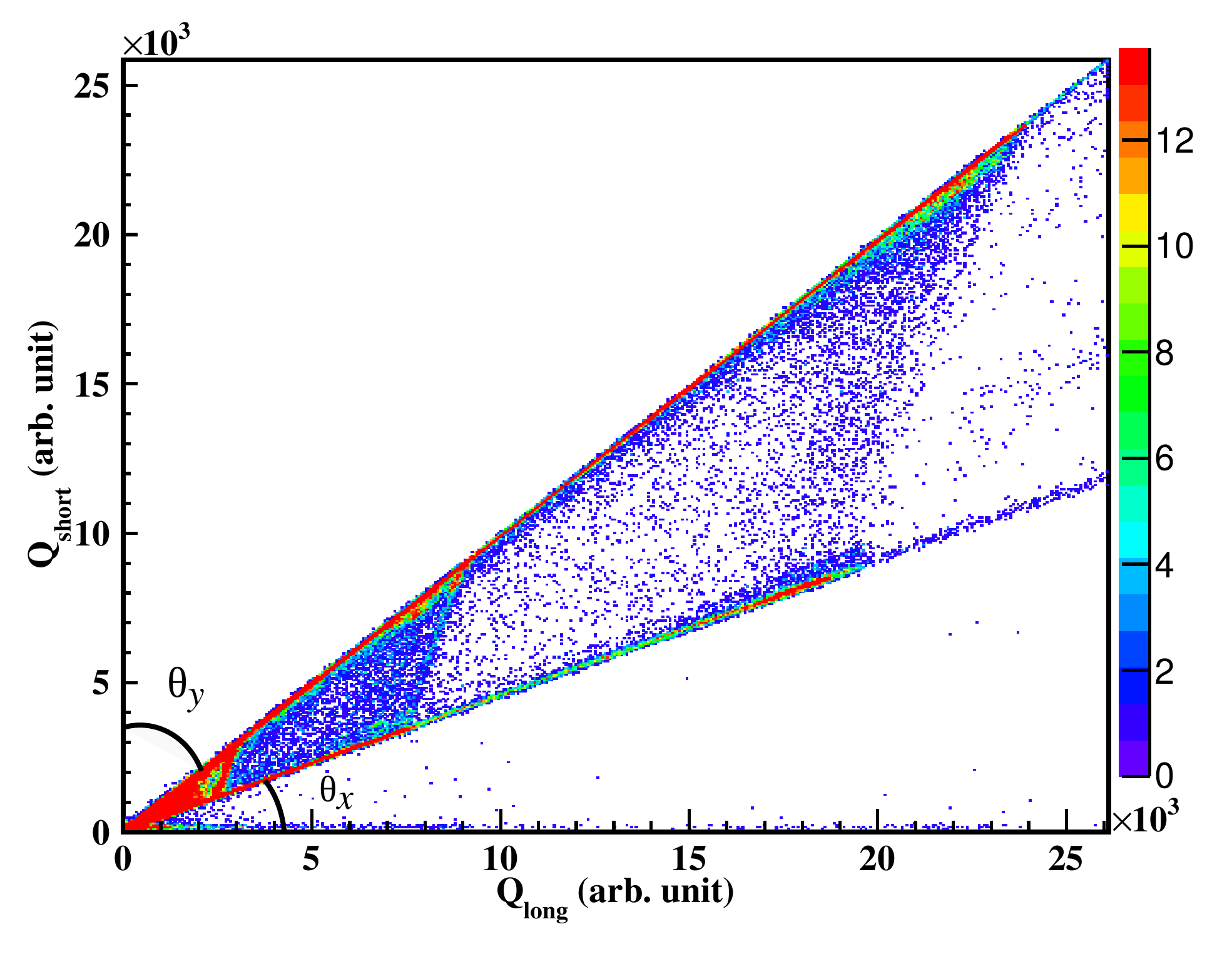}
\caption{\label{ECR2D}(Color online) A two dimensional histogram ($\rm Q_{short}$ vs $\rm Q_{long}$) constructed from the data acquired with DPP-PSD firmware in $^{11}B(p,\gamma)$ reaction at $E_p$~=~163~keV, for D2 detector. The angles $\theta_x$ and $\theta_y$ represent the tilt angles for pure events in NaI(Tl) and $\rm LaBr_3(Ce)$ crystals with respect to $\rm Q_{long}$ and $\rm Q_{short}$ axis, respectively.}
\end{figure}

The DPP-PSD firmware, enables the integration of the digitized pulse over two different pre-settable time windows, which are recorded together with the time stamp (in units of the sampling frequency) for every event. The recorded data is converted into ROOT~\cite{root} format for further processing. The short and long gate widths have been optimized to get the best energy resolution at 4.4~MeV and found to be 300~ns and 900~ns, respectively.
A typical 2D histogram of Q$_{\rm short}$ and Q$_{\rm long}$, integrated charges over short and long time gates, respectively (approximately corresponding to $E_{\rm LaBr_3}$ and $E_{\rm total}$) obtained in $^{11}B(p,\gamma)$ reaction at $E_p$~=~163~keV  is shown in Fig.~\ref{ECR2D} for the D2 detector. 
The $\rm LaBr_3(Ce)$ and NaI(Tl) spectra are constructed by appropriate projections of this spectrum. The projected spectra are analyzed using peak fitting routines in LAMPS and ROOT. 

 For studying the linearity of the energy response of the detector, the bias voltage for the PMT R7723-100 (V$_{max}$~=~--~2~kV) was varied from --~1.0~kV~to~--~1.8~kV.  It should be noted that at low PMT voltage, the signal from the NaI(Tl) was very poor and the threshold setting was limited by the 12~bit ADC. Further, in order to disentangle the detector and the PMT + base  (i.e., voltage divider circuit) set contribution, some measurements were repeated by interchanging these combinations, viz., D1 + PMT1, D2 + PMT2 and D1 + PMT2, D2 + PMT1. The effects due to temperature variations are expected to be negligible and not considered in the present study, since the ambient temperature was stable within $\pm1^\circ$C. The long term stability of the detectors was tested and found to be better than 1$\%$ over a period of 24 hours. 
\section{Linearity and Count Rate Effects}
\subsection{Linearity of Energy Response}
\begin{figure}[]
\centering
\includegraphics[scale=0.32]{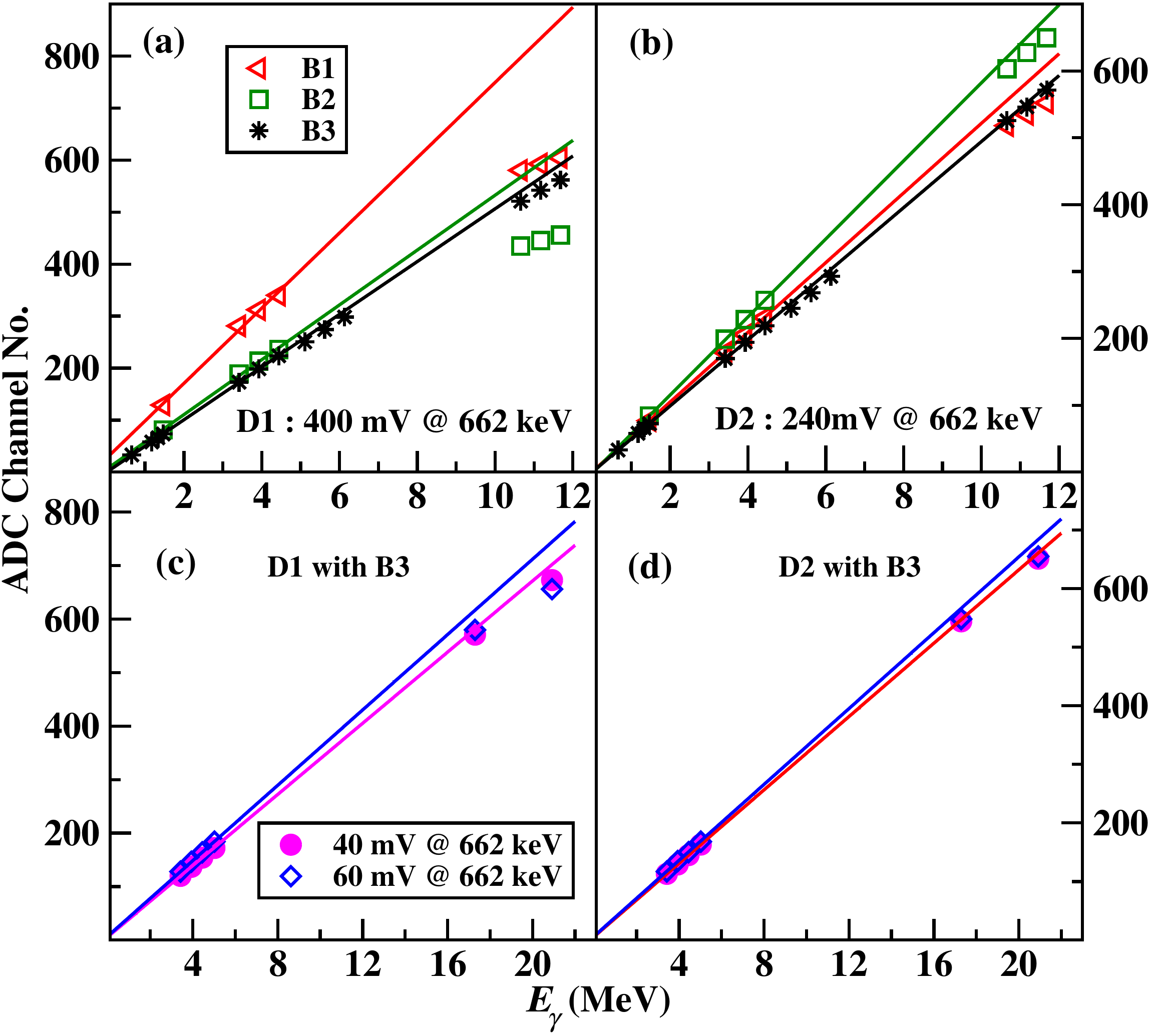}
\caption{\label{ecr}(Color online) Energy response linearity for $\rm LaBr_3(Ce)$ in phoswich detector : (a) for D1 with different voltage dividers upto $E_\gamma$~=~11.7~MeV; (b) same as (a) for D2; (c) for D1 with B3 voltage divider and upto $E_\gamma$~=~22.6~MeV at two different applied voltages, (d) same as (c) for D2. The symbols represent the measured data, while the straight lines are the extrapolation of linear calibration using gamma ray energies up to 4.4~MeV.}
\end{figure}

As mentioned in the earlier section, one of the major concerns for $\rm LaBr_3(Ce)$ detector is the non-linearity of the detector response at high energy owing to its high light yield. A total of three types of the voltage divider circuits were used in the measurements, namely, B1 -- the standard Hamamatsu make resistive voltage divider E5859-15, B2 -- a simple modification from the B1 with the gain of first two dynode stages reduced to half of the original values and B3 -- a passive voltage divider developed by IPHC~Strasbourg group for the PARIS detector~\cite{strasbourg}. 
The standard configuration for the dividers uses charge pump capacitors only at the last few gain stages. However, in order to handle larger signals arising from the $\rm LaBr_3(Ce)$ elements and higher counting rates, the modified divider uses charge pump capacitors at all stages with higher value capacitors than in standard configurations.
The raw pulse heights from the two detectors are different, mainly due to PMT + voltage divider combination. Initially, for comparative study of three voltage dividers, the same high voltage was applied and hence the pulse heights were different for different voltage dividers.
For linearity and count rate studies  with B3 voltage divider, the high voltage was adjusted to get the similar pulse height from the two detectors to ensure similar PMT chain current. The detector signals were suitably attenuated using a passive attenuator to make it acceptable for the ADC input range and to adjust the full energy scale to $\sim$~17~MeV and 25~MeV for the radiative capture reactions at $E_p$ = 163~keV and 7.2~MeV, respectively. The energy responses of the detectors were found to be linear upto 4.4~MeV for all the voltage dividers upto the bias voltage of $\sim$~1.8~kV. Fig.~\ref{ecr}(a)~-~(b) show the comparison of energy response with different voltage dividers upto $E_\gamma$~=~11.7~MeV for D1 and D2, respectively. Non-linearity ($\alpha$) of the energy response is estimated as the percentage deviation from the linear extrapolation of energy calibration upto 4.4~MeV. Table~\ref{ECRNonLinear} gives the measured non-linearity for different voltage dividers with same high voltage together with the anode pulse height (A1) for 662~keV gamma ray. The detector D1 having larger pulse height, shows higher non-linearity as expected. 

\begin{table}[H]
\centering
\caption{The non-linearity ($\alpha$) of the $\rm LaBr_3(Ce)$ detector with different voltage dividers (B1, B2, B3) at $E_\gamma$ = 11.7~MeV. The anode pulse height (A1) for 662~keV in the $\rm LaBr_3(Ce)$ crystal is also listed ($\rm V_{D1}$ = --~1.6~kV and $\rm V_{D2}$ = --~1.8~kV).}
\label{ECRNonLinear}
\begin{tabular}{|c|c|c|c|c|c|c|}
\hline
& \multicolumn{2}{ |c| }{\bf{B1}}& \multicolumn{2}{ |c| }{\bf{B2}} & \multicolumn{2}{ |c| }{\bf{B3}} \\ \cline{2-7}
\bf{Detector} & \multicolumn{1}{ |c| }{A1} & \multicolumn{1}{|c|}{$\alpha$} & \multicolumn{1}{ |c| }{A1} & \multicolumn{1}{|c|}{$\alpha$} & \multicolumn{1}{ |c| }{A1} & \multicolumn{1}{|c|}{$\alpha$} \\ 
& \multicolumn{1}{ |c| }{(V)} & \multicolumn{1}{|c|}{(\%)} & \multicolumn{1}{ |c| }{(V)} & \multicolumn{1}{|c|}{(\%)} & \multicolumn{1}{ |c| }{(V)} & \multicolumn{1}{|c|}{(\%)} \\ \cline{1-7}
\bf{D1} & \multicolumn{1}{ |c| }{1.5}& \multicolumn{1}{ |c| }{31} & \multicolumn{1}{ |c| }{1.3}  & \multicolumn{1}{|c|}{27} & \multicolumn{1}{ |c| }{0.4}  & \multicolumn{1}{|c|}{5} \\ \cline{1-7}
\bf{D2} & \multicolumn{1}{ |c| }{0.68}& \multicolumn{1}{ |c| }{9} & \multicolumn{1}{ |c| }{0.6}  & \multicolumn{1}{|c|}{5} & \multicolumn{1}{ |c| }{0.24}  & \multicolumn{1}{|c|}{1} \\ \cline{1-7}
\end{tabular}
\end{table}

It is evident that the B3 type voltage divider is significantly better even at $E_\gamma\sim$~12~MeV. Therefore, linearity upto 22.6 MeV was investigated only for B3 type divider for A1~$\sim$~40~mV ($\rm V_{D1}$~=~--~1.040~kV and $\rm V_{D2}$~=~--~1.310~kV) and 60 mV ($\rm V_{D1}$~=~--~1.123~kV and $\rm V_{D2}$~=~--~1.418~kV). The energy range ($E_{\rm full-scale}$) was adjusted to $\sim$ 30~MeV with suitable attenuation. The results are shown in  Fig.~\ref{ecr}~(c)~-~(d) and tabulated in Table~\ref{PELLNonLinear}. It should be mentioned that the centroid for 18.1 and 22.6~MeV gamma rays were obtained from the simulated spectra after folding the resolution function, as described in the later section.

 For both detectors, at A1~$\sim$~40~mV setting the non-linearity is about 2--4\% at $E_{\gamma}$ =~22.6 MeV. However, it should be noted that the bias voltage is significantly lower ($\sim$~--1.0~to~--1.3~kV), which impacts both energy and time resolution. Hence A1~$\sim$~60~mV, which yields $\alpha$~$\sim$~10\%, would be an optimum choice. In case of NaI(Tl), the observed non-linearity in all of the above configurations was smaller than 1\% upto $E_\gamma\sim$ 11.7~MeV.

\begin{table}[]
\centering
\caption{The non-linearity ($\alpha$) of the $\rm LaBr_3(Ce)$ detector with B3 type voltage divider at $E_\gamma$ = 18.1 and 22.6~MeV for different high voltages (A1) has same meaning as in Table~3.}
\label{PELLNonLinear}
\begin{tabular}{|c|c|c|c|c|}
\hline
 & \multicolumn{4}{|c|}{\bf{Non-linearity (\%)}} \\ \cline{2-5}
 & \multicolumn{2}{|c|}{$E_\gamma$ = 18.1~MeV} & \multicolumn{2}{|c|}{$E_\gamma$ = 22.6~MeV} \\ \cline{2-5}
\bf{Detector}  & \multicolumn{1}{|c|}{A1=60} & \multicolumn{1}{|c|}{A1=40} & \multicolumn{1}{|c|}{A1=60} & \multicolumn{1}{|c|}{A1=40} \\ 
 & \multicolumn{1}{|c|}{(mV)} & \multicolumn{1}{|c|}{(mV)} & \multicolumn{1}{|c|}{(mV)} & \multicolumn{1}{|c|}{(mV)} \\ \cline{1-5}
\bf{D1} & \multicolumn{1}{|c|}{6} & \multicolumn{1}{|c|}{2} & \multicolumn{1}{|c|}{12} & \multicolumn{1}{|c|}{4} \\ \cline{1-5}
\bf{D2} & \multicolumn{1}{|c|}{3} & \multicolumn{1}{|c|}{1} & \multicolumn{1}{|c|}{4} & \multicolumn{1}{|c|}{2} \\ \cline{1-5}
\end{tabular}
\end{table}

\subsection{Detector Gain Stability as a Function of Count Rate}
Considering excellent time resolution of the $\rm LaBr_3(Ce)$ detector, it is envisaged that the PARIS detector will be used in a close geometry thereby enhancing the detection efficiency. Thus, high count rate handling capability is essential. Generally, during in-beam experiment any variations in beam intensity can lead to fluctuations in count rate. Given the fact that the phoswich detector pulse has a  width of about $\sim$~1~$\mu$s, in principle the detector has a large count rate handling capacity ($\sim$~MHz). However, if the phoswich detector + PMT + voltage divider assembly is not robust with respect to the count rate variations, it will result in the degradation of the energy resolution. At high count rate, a combination of pile-up and variation in the PMT gain comes into effect and the peak shift will depend on the dominance of these two effects.

\begin{figure}[h]
\centering
\includegraphics[scale=0.305]{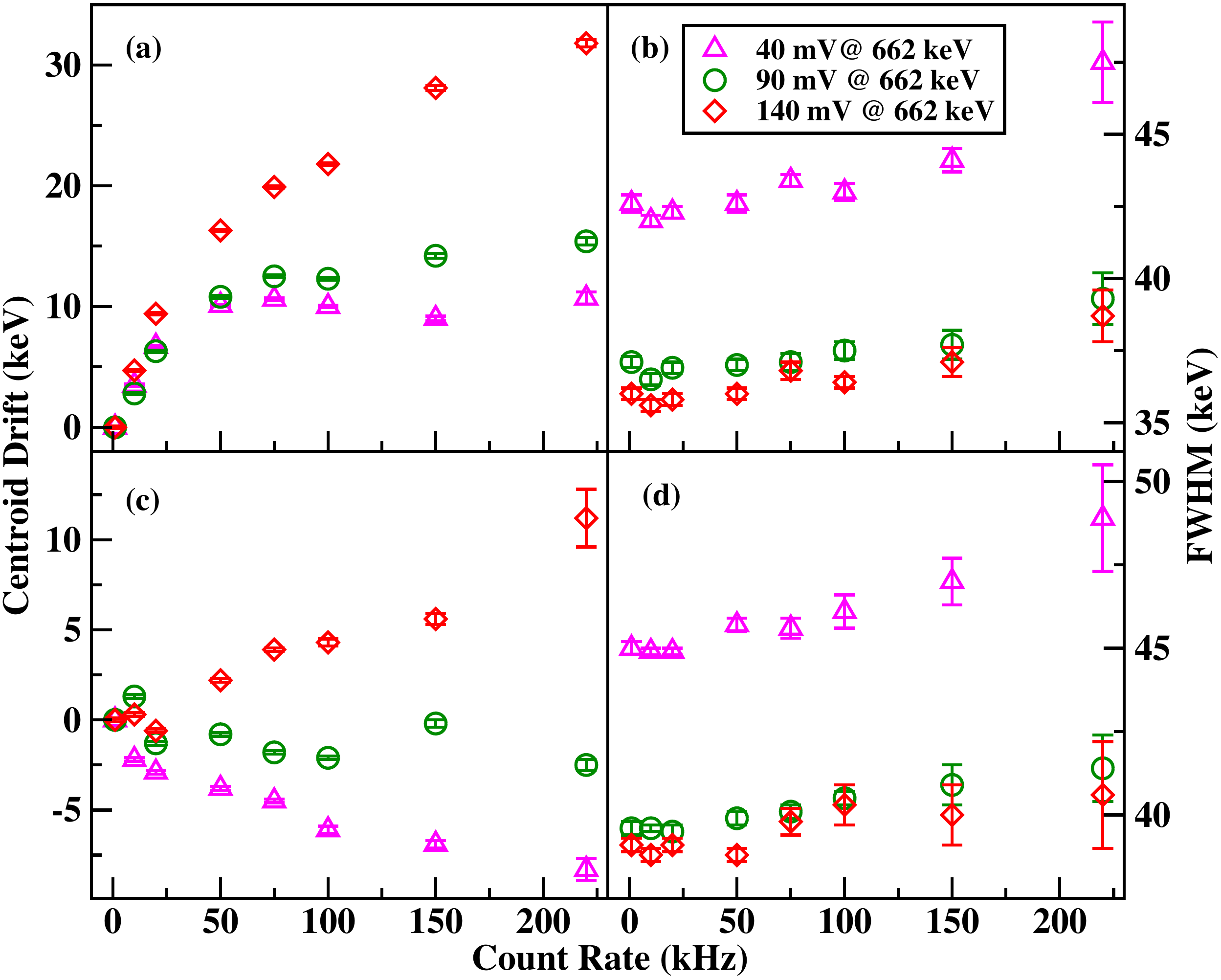}
\caption{\label{D1count}(Color online) 
Variation of the centroid shift and FWHM of 835 keV gamma ray peak in $\rm LaBr_3(Ce)$ as a function of count rate  (a) and (b) for D1 detector;  (c) and (d) for D2 detector. Measurements were done using B3 type voltage divider for different high voltages and corresponding pulse heights (A1) for 662 keV are listed in legends.}
\end{figure}

To estimate the robustness of phoswich detector + PMT for count rate variation, the centroid drift and the resolution of 835~keV gamma ray (from $\rm ^{54}Mn$ source) in the $\rm LaBr_3(Ce)$ were studied over a broad  count rate range of 2~-~220~kHz. 
This was achieved by keeping a weak $\rm ^{54}Mn$ source (8.3~mBq) at a fixed distance from the detector and by varying the position of a strong  $\rm ^{137}Cs$  source (4~MBq). The choice of the low energy ($E_\gamma$~=~662~keV) strong source ensured that the centroid of 835~keV gamma ray line was minimally affected due to pile-up, summing and Compton edge from the strong source. A stabilization time of $\sim$~30 min was allowed for each setting. The drift in centroid for each detector was measured with respect to the lowest count rate (2~kHz).
 Fig.~\ref{D1count} shows variation in peak centroid and FWHM for D1 (panels (a)~-~(b)) and for  D2 (panels (c)~-~(d)) with both PMT bases having the B3 type divider. Data correspond to different high voltages and the respective anode pulse heights (A1) for 662~keV gamma ray are mentioned in the legends.
 
At the lowest voltage (A1~=~40~mV), the centroid shift for D1 is less than 1.5\% (11~keV) even at the highest count rate of 220~kHz, while the FHWM worsens by $\sim$~13\% ($\sim$~5~keV). At the highest count rate, errors in FWHM are larger due to fitting errors. The shifts at high count rate are expectedly worse (centroid shift $\sim$~5\%) for higher pulse heights (A1~=~140~mV). It may be noted that overall the shifts are smaller for D2 as compared to the D1.
 As it is evident from the Fig.~\ref{D1count}, the shifts in the centroid of  D1 and D2 are in the opposite direction. By interchanging the PMT + base sets on two detectors (i.e. with D1 + PMT2 and D2 + PMT1), it was verified that the count rate effects are intrinsic to the PMT + voltage divider combination. It should be mentioned that larger count rate variation at 220~kHz (centroid shift $\sim$3.7\% i.e. 31~keV for A1~=~40~mV) was observed with Wiener EHSF020n (16~- channel, 2~kV, 4~mA) high voltage supply. It is clear that the performance of the detectors is quite stable in the range 50 to 150 kHz in the region of interest (A1~=~40 to 140~mV).

\begin{figure}[h]
\centering
\includegraphics[scale=0.31]{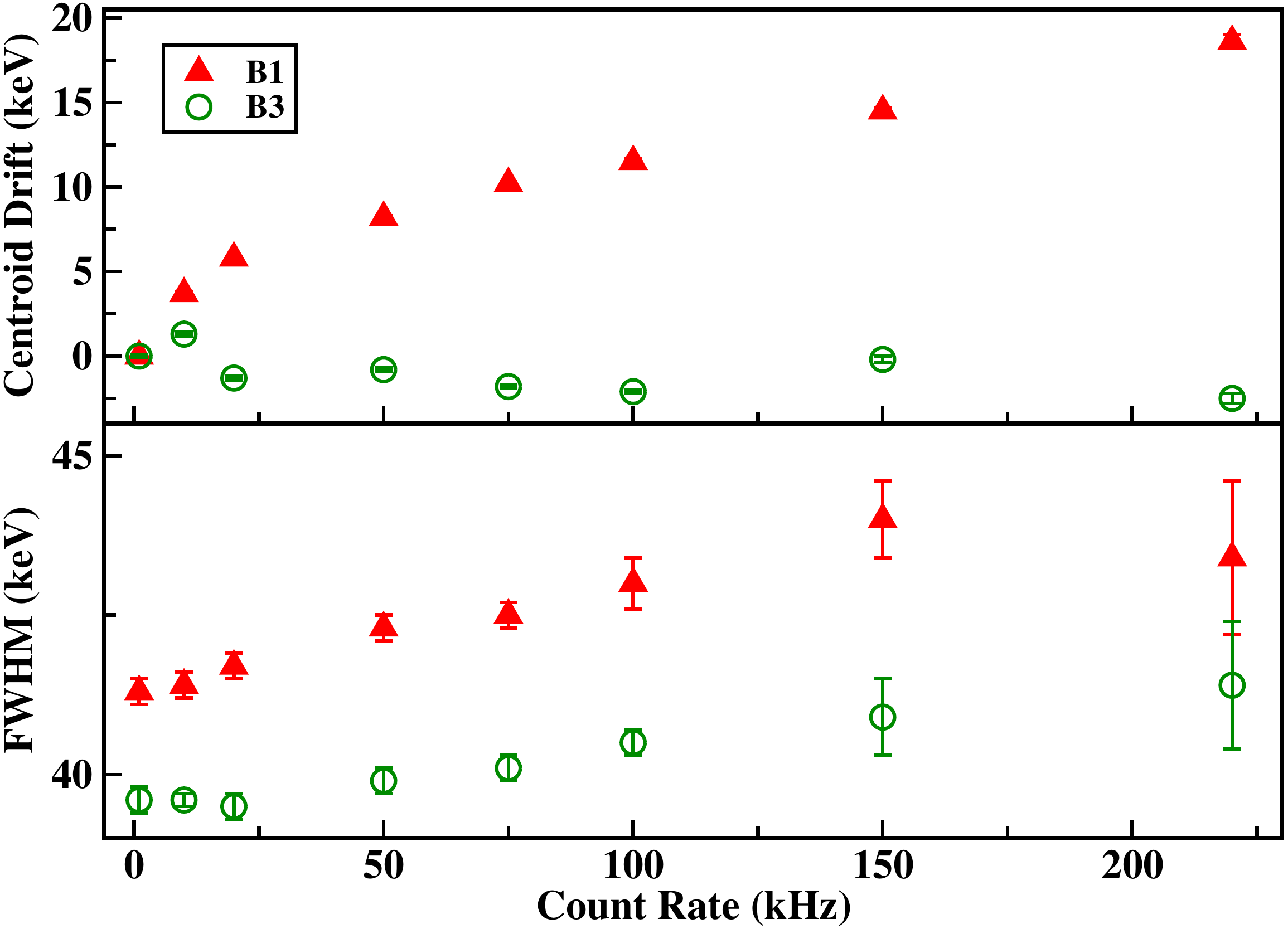}
\caption{\label{STDvsSTRS}(Color online) Variation of the centroid shift and FWHM of 835 keV gamma ray peak as a function of count rate for D2 detector with B1 and B3 type voltage dividers for pulse height at 662 keV, A1~=~90~mV.}
\end{figure}

The count rate handling capability was also tested with the standard base B1 (90~mV @ 662~keV) and results are shown in Fig.~\ref{STDvsSTRS}. It can be seen that even at 10~kHz, the centroid shift ($\sim$~3.7~keV)  with  the B1 voltage divider is significantly larger than that for the B3 ($\sim$~1.3~keV). A similar behavior is seen for the FWHM.  Hence, it is evident that performance of the B3 voltage divider is better in terms of linearity and stability against count rate variation.

\section{Analysis of Detector Response}

 The phoswich detector response is simulated using GEANT4.10~\cite{geant4}. 
 Detector geometry as per design is employed including 1~mm thick Aluminum cover but the optical window is not considered. The source geometry is taken to be point like for reaction measurements and most of the sources. In the case of a sealed Am-Be source, the SS capsule is incorporated in the simulation and the source is assumed to be uniformly distributed within the capsule. The Aluminum vacuum chamber and the Lead absorber are included for high energy gamma rays. Typically, 10$^8$ events are generated for each simulation set. The simulated spectra have been analyzed in the ROOT in a manner identical to experimental data. 
 
In the present work, energy resolution is studied separately for both LaBr$_3$(Ce) and NaI(Tl) segments. Given the nature of phoswich detector, majority of the low energy gamma rays will have primary interaction in the LaBr$_3$(Ce) segment.  
Hence, efficiency of the LaBr$_3$(Ce) is measured for low energy gamma rays at a distance of 10--20~cm, corresponding to a typical PARIS cluster configuration and compared with simulations.
The energy resolution and efficiency at high energies are deduced from a comparison of the experimental spectra with simulated spectra. 
  
\begin{figure}[]
\centering
\includegraphics[scale=0.315]{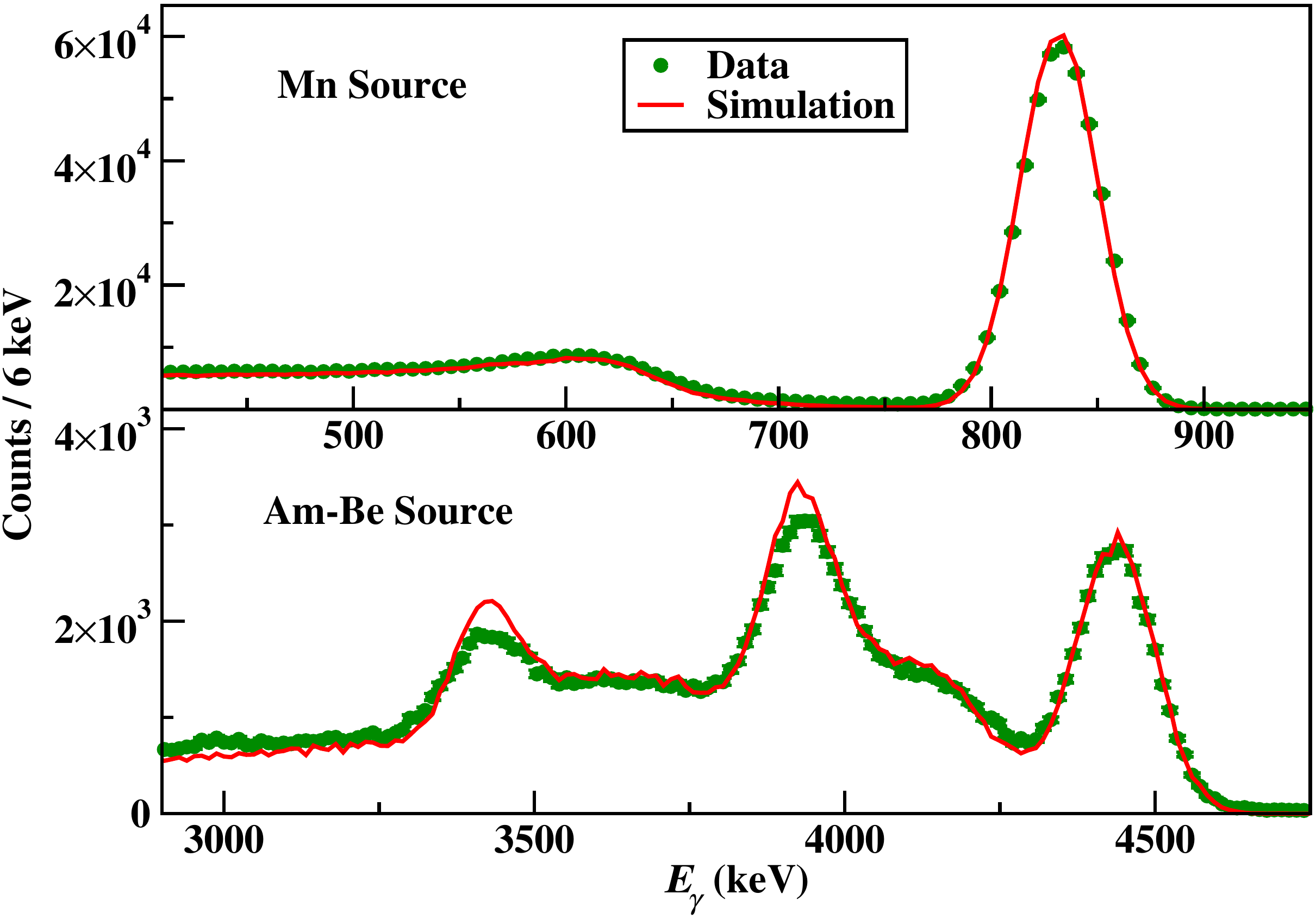}
\caption{\label{sim1} A Comparison of simulated (after incorporating the resolution) and experimental spectra for 835 keV and 4.4 MeV without any normalisation.}
\end{figure}

A comparison of the simulated and the background corrected experimental spectra for 835 keV and 4.4 MeV gamma rays for D1-LaBr$_3$(Ce) detector is shown in Fig.~\ref{sim1} on an absolute scale, after folding the resolution of 5.1\% for 835~keV and 3.3\% for 4.4~MeV. The excellent agreement between simulations and data for complete line shape including escape peaks for the 4.4 MeV is evident.

\subsection{Energy Resolution}
 For low energy gamma rays ($E_\gamma$~<~4.4~MeV), the energy resolution for LaBr$_3$(Ce) was obtained by a simple gaussian fit to the observed photo-peak. This was then fitted to the following equation~\cite{labr3}
\begin{equation}\label{ResFunc}
FWHM/E_\gamma(\%) = \sqrt{a+b/E_\gamma}.
\end{equation}

\begin{figure}[h]
\centering
\includegraphics[scale=0.32]{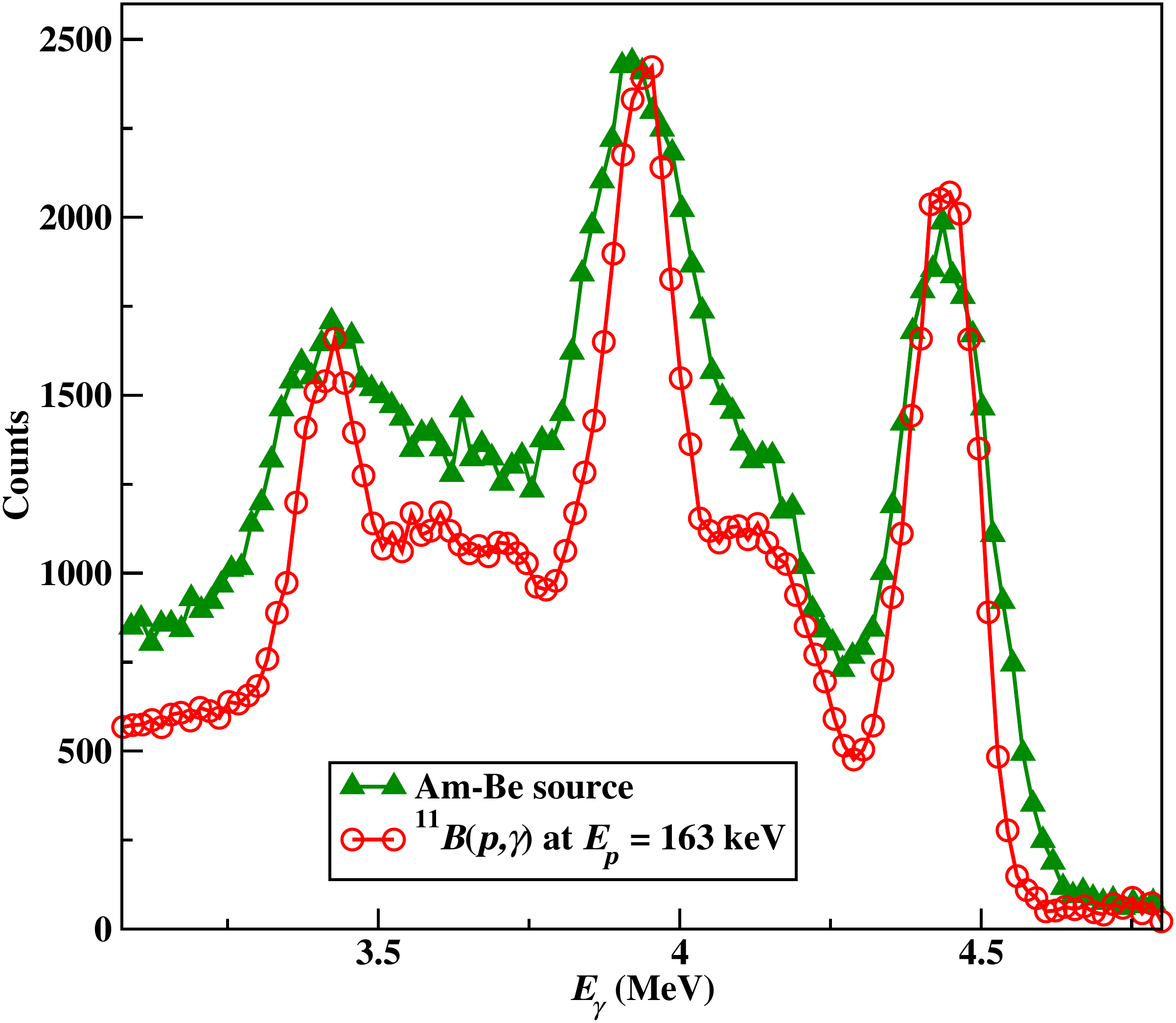}
\caption{\label{SourceVsBeam}(Color online) Measured spectrum in D1-$\rm LaBr_3(Ce)$  for 4.4 MeV gamma ray from Am-Be source (FWHM~$\sim$~3.6\%) and from $^{11}B(p,\gamma)$ reaction at $E_p$ =~163~keV (FWHM~$\sim$~2.7\%).}
\end{figure}

The simulated data is folded with a Gaussian resolution function, where the FWHM is given by Eq. (4.1). Then both the parameters ($a$, $b$) were varied to reproduce the experimentally observed lineshape over a wider energy range (662~keV - 22.6~MeV) for each detector. Doppler correction for $E_p$~=~7.2 MeV (relative recoil velocity, $\beta_{recoil}$ = 0.01) was incorporated in the simulations for gamma rays emitted in the reaction, while that for $E_p$~=~163 keV is neglected as $\beta_{recoil}$ = 0.002 is very small.
Figure~\ref{SourceVsBeam} shows the spectrum for 4.4 MeV gamma ray from the source and from the $^{11}B(p,\gamma)$ reaction at $E_p$ = 163 keV. The effect due to Doppler broadening in the Am-Be source spectrum~\cite{mowlavi} is clearly visible and also re-emphasizes excellent energy resolution of the $\rm LaBr_3(Ce)$, even in the phoswich configuration.

\begin{figure}[]
\centering
\includegraphics[scale=0.315]{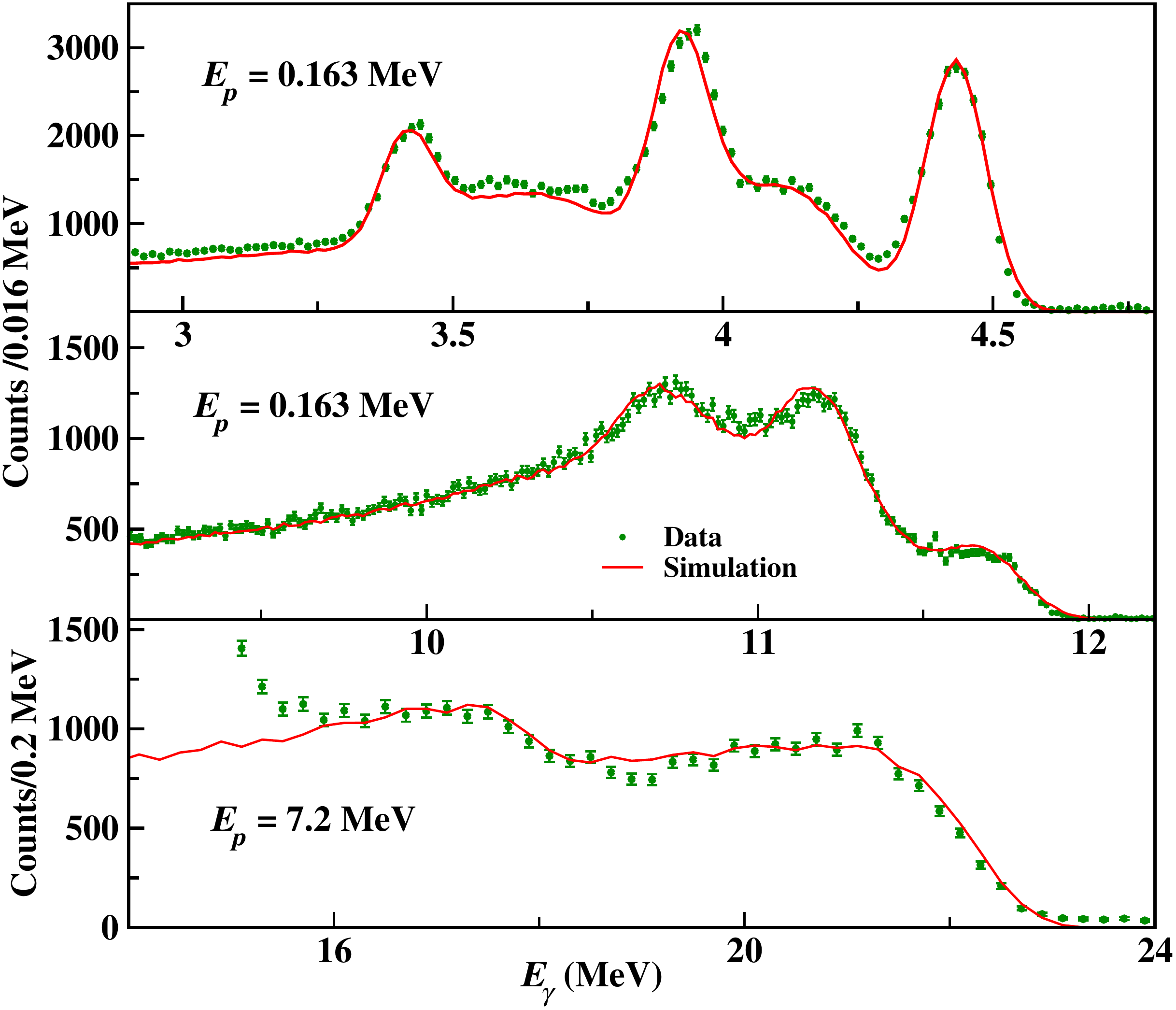}
\caption{\label{simulation}(Color online) Experimental spectra (solid circle) for different gamma rays together with the simulated spectrum (line) after folding in resolution function (see text).}
\end{figure}

Figure~\ref{simulation} shows the spectra of 4.4 MeV, 11.7 MeV, 18.1 MeV and 22.6~MeV gamma rays from $^{11}B(p,\gamma)$ reaction together with simulated spectra (after folding with resolution function). Since the proton charge could not be measured, simulated spectra are normalized to area under the the photo-peak for $E_\gamma$~=~4.4~MeV.
For high energy gamma rays (viz., 11.7 and 22.6~MeV), spectra are normalized to the integral counts in energy range covering the 2$^{nd}$ escape-peak to the photo-peak.
Relative intensity of $E_\gamma$~=~18.1 MeV with respect to $E_\gamma$~=~22.6 MeV is taken to be 0.24 for the best fit to the data (Ref. value is 0.25~\cite{snover}).
It can be seen that in the low energy proton reaction ($E_p$~=~163 keV), where no additional background is present, excellent agreement is observed between simulation and data. In the high energy proton induced reaction ($E_p$~=~7.2 MeV), the peak shapes for 18.1 and 22.6~MeV gamma rays are well reproduced but in the experimental spectra there is an excess left tail. This could be due to beam induced gamma/neutron background and neutron capture reactions in Lanthanum, Bromine and Iodine.

\begin{figure}[h]
\centering
\includegraphics[scale=0.32]{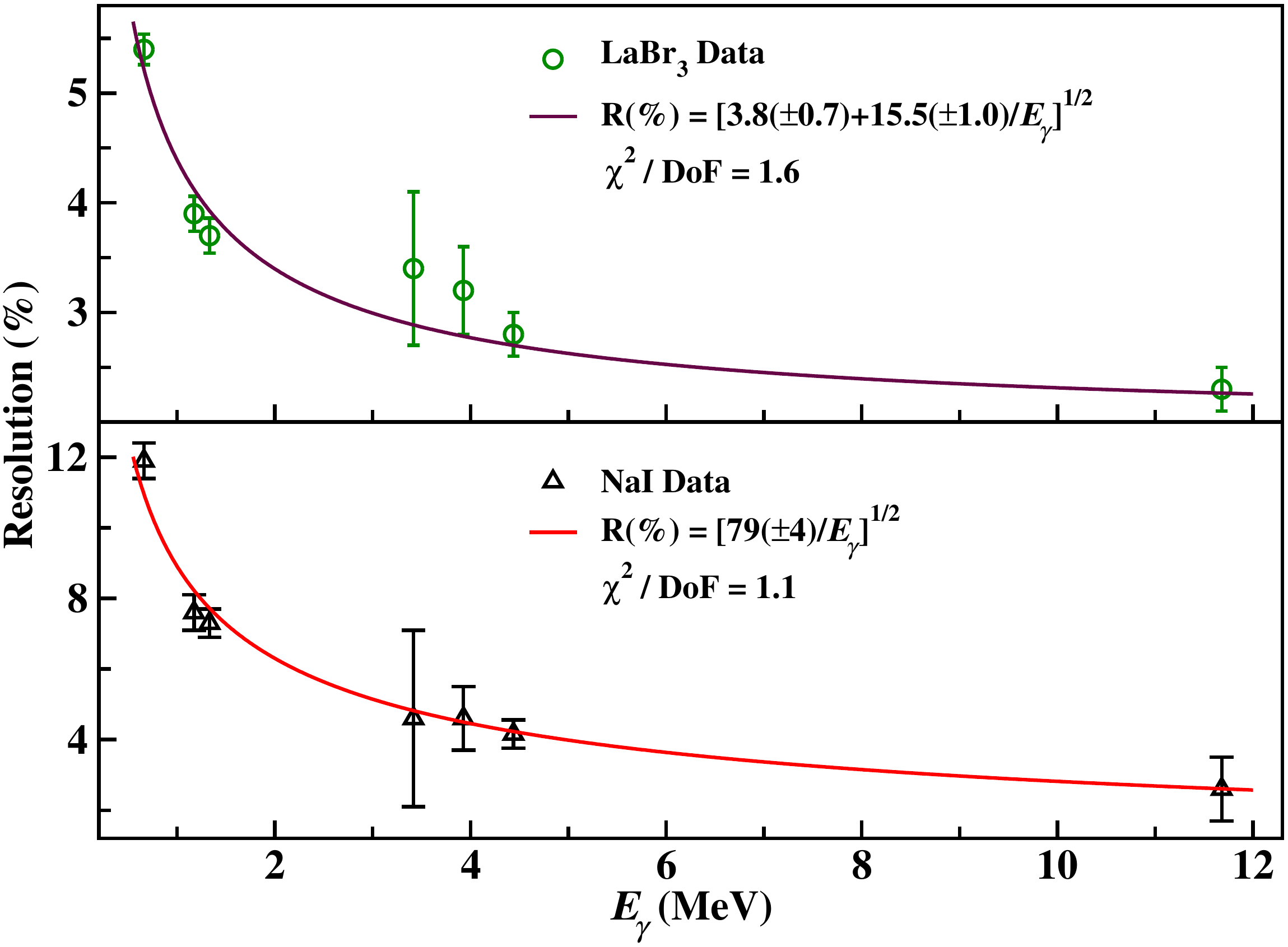}
\caption{\label{eresol}(Color online) Energy resolution of $\rm LaBr_3(Ce)$ (top) and NaI(Tl) (bottom) crystal of D2 detector up to 12~MeV (A1=~60~mV). The symbols represent the measured resolution and the solid lines represent the fitted function.}
\end{figure}

Figure~\ref{eresol} shows the energy resolution of the $\rm LaBr_3(Ce)$ segment (top panel) as a function of gamma ray energy from 0.6 to 12~MeV for the D2 detector (A1~=~60 mV). The symbols correspond to the data and the lines represent the fit function (Eq.~(\ref{ResFunc})). 
 The observed energy resolution of D1 (D2) detector  shows  a small improvement  from 2.9 (2.8)\%  to  2.5 (2.3)\% for $E_\gamma$~=~4.4 and 11.7~MeV, respectively. The energy resolution for NaI segment, which could be fitted to a simple $1/\sqrt{E_\gamma}$ function, is shown in the bottom panel. 
 In the case of $E_p$~=~7.2~MeV induced reaction, the PMT was operated at lower voltages (A1~=~40 mV) to minimize non-linearity. As a result, low energy peaks  (662~-~1332~keV) could not be cleanly measured. Further, the 4.4 MeV (from both Carbon and Boron) and 5.02~MeV gamma rays (from Boron) cannot be separated. Hence, the fit to resolution function was obtained using fewer points. In this configuration, the energy resolution obtained for both detectors was around $\sim2.1\%$  at 22.6 MeV. 

\subsection{Time Resolution}
\begin{figure}[]
\centering
\includegraphics[scale=0.37]{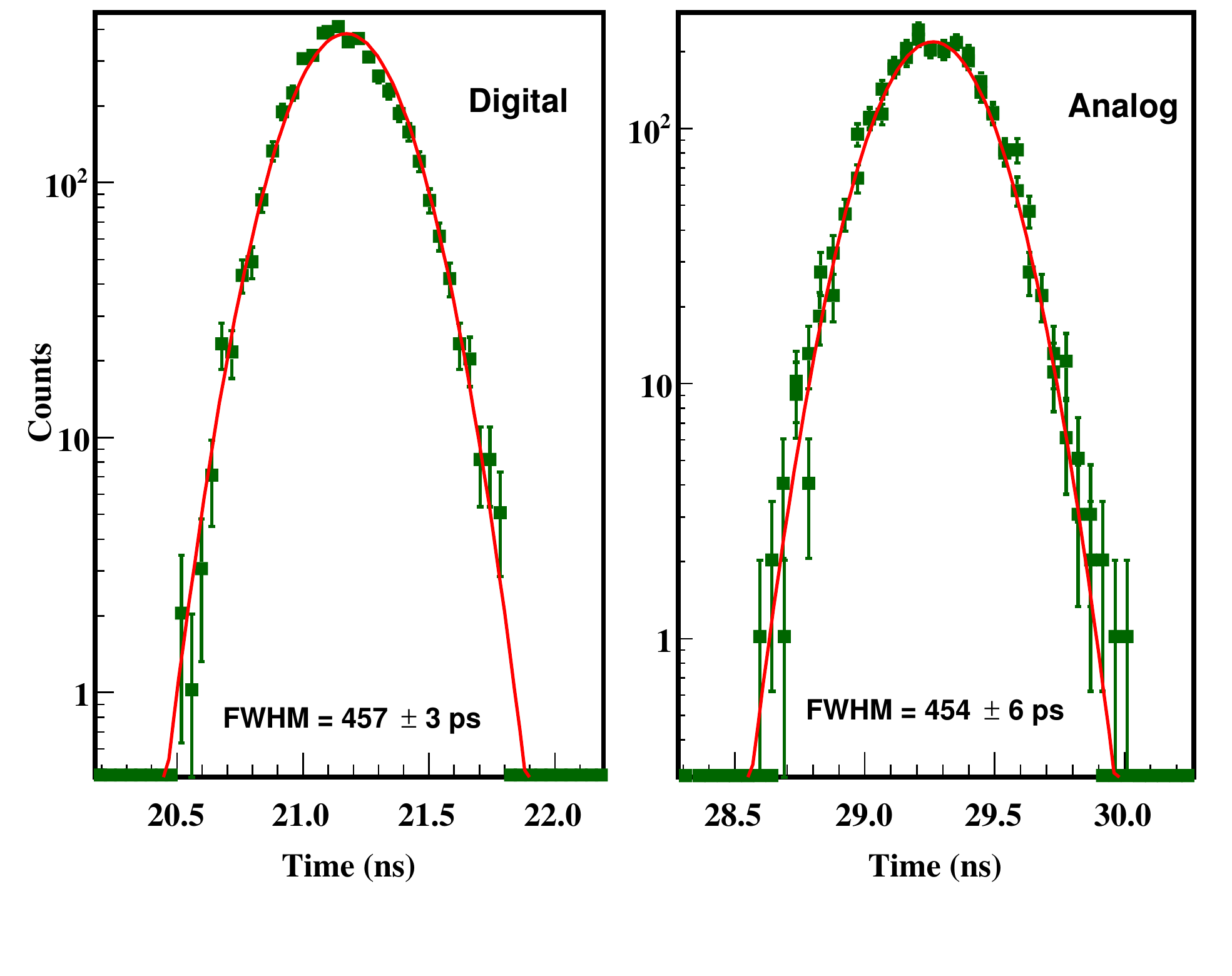}
\caption{\label{resol}(Color online) Energy gated time-of-flight spectra of two $\rm LaBr_3(Ce)$ crystals of D1 and D2 using $\rm ^{60}Co$ source with V1751 (1~GS/s) digitizer (left panel) and with conventional analog electronics (right panel). The red line represents the gaussian fit to the experimental data.}
\end{figure}

To extract timing information with a better accuracy, an algorithm implementing constant fraction discrimination with delay of 6~ns ($\sim$~67\% of the rise-time) and 20\% fraction has been developed and incorporated in the online WAVEDUMP acquisition software~\cite{cfd}. Linear interpolation was used to calculate the zero-crossing of the CFD pulse. Measurements were carried out with detectors D1 ($\rm V_{D1}$~=~--~1.6~kV) and D2 ($\rm V_{D2}$~=~--~1.8~kV ) placed at 180 degree facing each other, with the  $\rm ^{60}Co$ source mounted at the center at $\sim$~2~cm from either detector. The detector pulses were sampled with V1751 (1~GS/s) digitizer and $\rm Q_{short}$, $\rm Q_{long}$ and $\rm T_{fine}$ (obtained with CFD) were stored. The left panel of Fig.~\ref{resol} shows the energy gated time-of-flight (TOF) spectra of two detectors (i.e. 1173 or 1332~keV photo-peak in either LaBr$_3$(Ce) crystals). The right panel shows the time spectrum obtained using the conventional analog electronics consisting of a CFD and TAC. It can be seen that FWHM~(digital)~$\sim$ 457~$\rm \pm$~3~ps is similar to FWHM~(analog)~$\sim$ 454~$\rm \pm$~6~ps. Using an additional $\rm CeBr_3$ detector, the time resolution of individual $\rm LaBr_3(Ce)$ crystal was measured to be FWHM (D1)~$\sim$~323~ps and FWHM (D2)~$\sim$~309~ps.

It should be noted that the above configuration corresponds to a dynamic energy scale of $\sim$~2~MeV in both detectors. The time resolution of the detectors was also measured at lower operating voltages (A1~=~60~mV) and was found to worsen by about 12$\%$ (FWHM~$\sim513~\rm \pm$~3~ps), which is still excellent for the neutron-$\gamma$ separation by TOF measurement at close distances.

\subsection{Efficiency}
\begin{figure}[]
\centering
\includegraphics[scale=0.32]{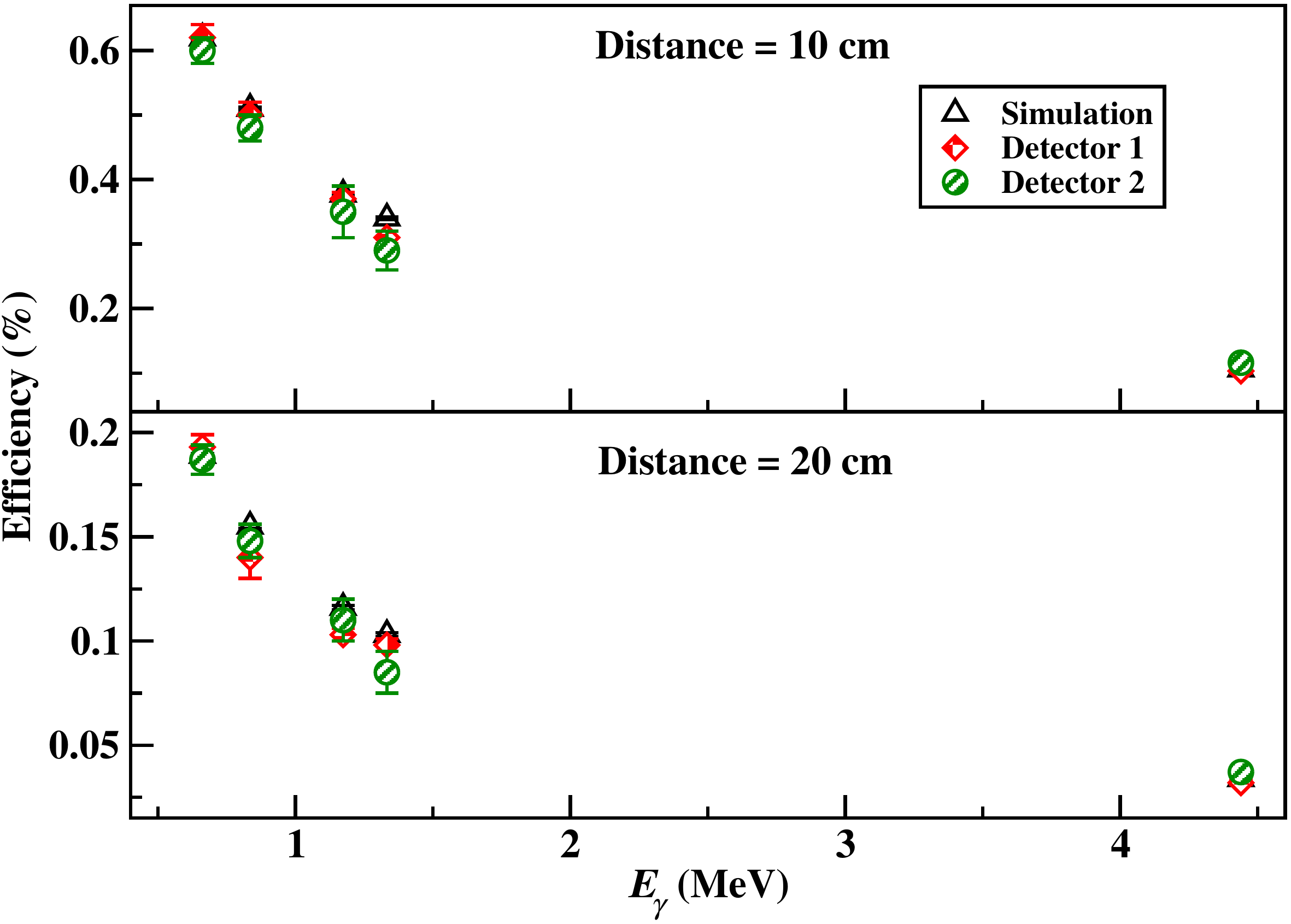}
\caption{\label{ppeff}(Color online) The measured and simulated photo-peak efficiency of the $\rm LaBr_3(Ce)$ crystals of D1 and D2 at a distance of 10 and 20~cm from the face of the detector.}
\end{figure}

The photo-peak efficiencies of the $\rm LaBr_3(Ce)$ crystal of D1 and D2 were measured for $E_\gamma$~=~662~keV to 4.4~MeV at varying distances of 10, 15 and 20~cm from the face of the detector. The results are shown in Fig.~\ref{ppeff} along with the simulations for 10 and 20~cm. The measured data agree with simulations at all distances.
The efficiency of $\rm LaBr_3(Ce)$ crystal was also measured by keeping the source at a distance of 10~cm from the side of the detector and were found to be consistent. As mentioned earlier the intrinsic efficiency at high energy could not be extracted. However, the 4.4~MeV and 11.7~MeV gamma rays are emitted in coincidence in $ ^{11}B(p,\gamma)$ reaction at $E_p$~=~163~keV. Hence, the relative intrinsic efficiency of 11.7~MeV with respect to 4.4~MeV could be measured and was found to be $\sim$~24\%.

\subsection{Total Energy Spectrum} 
The construction of total energy ($\rm E_{tot}$) spectrum employing the add-back of energy deposited in $\rm LaBr_3(Ce)$ and NaI(Tl) crystals was demonstrated by M.~Zi\k{e}bli{\'{n}}ski {\it et. al.}~\cite{paris1}. The formalism used to construct the $E_{\rm tot}$  in the present case is described in the following. As mentioned earlier, $\rm Q_{short}$ and $\rm Q_{long}$ represent the  integrated charge over the  short and long time gates, respectively. For a given event, let $E_1$ and $E_2$ be the energy deposited in the individual LaBr$\rm _3(Ce)$ and NaI(Tl) phoswich elements, respectively and the $\rm q_1(\it E_1)$ and $\rm q_2(\it E_2)$ be the corresponding intrinsic charges. 

The recorded charges can be represented as 
\begin{eqnarray}
\rm Q_{short}&=& \rm q_1(\it E_1)\rm cos\theta_y~+~\rm q_2(\it E_2)\rm sin\theta_x\\ 
\rm Q_{long}&=& \rm q_1(\it E_1)\rm sin\theta_y~+~\rm q_2(\it E_2)\rm cos\theta_x,
\end{eqnarray}
where $\theta_x$ and $\theta_y$ represent the tilt angles for pure events in NaI(Tl) and $\rm LaBr_3(Ce)$ crystals with respect to $\rm Q_{long}$ and $\rm Q_{short}$ axis, respectively, as shown in  Fig.~\ref{ECR2D}. In practice, the tilt angles are defined for the centroid of the NaI(Tl) and $\rm LaBr_3(Ce)$ bands.
The intrinsic charges can be extracted as 
\begin{eqnarray}
\rm q_1(\it E_1)&=& k \times (\rm Q_{short}cos\theta_x- \rm Q_{long}sin\theta_x )\\
\rm q_2({\it E_2})&=& k \times (-\rm Q_{short}sin\theta_y+ \rm Q_{long}cos\theta_y ),
\end{eqnarray}
with
\begin{equation}
k=\frac{1}{\rm cos (\theta_y +\theta_x)}.
\nonumber
\end{equation}

\begin{figure}[H]
\centering
\includegraphics[scale=0.42]{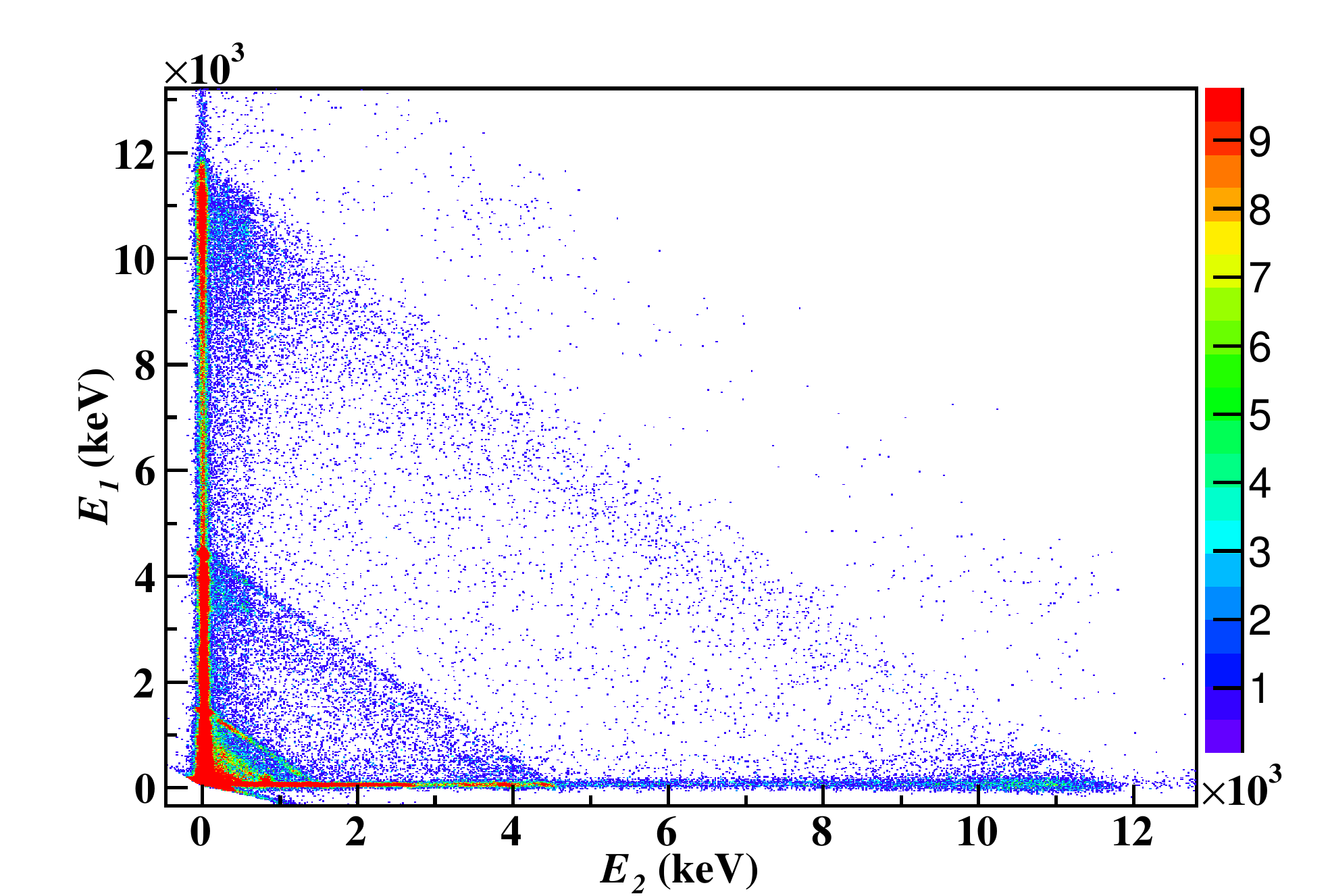}
\caption{\label{addback1}(Color online) Two dimensional spectrum of Fig.~3 after transformation as described in Eq.~(4.4)--(4.6) with $\theta_x$~=~24$^\circ$ and $\theta_y$~=~45$^\circ$, for D2 detector with A1~=~60~mV.}
\end{figure}

Using gamma rays listed in Table~\ref{source} and~\ref{reaction}, the calibrations for $E_1$ vs q$_1$ and $E_2$ vs q$_2$ are obtained. Figure~\ref{addback1} shows the 2D spectrum of $E_1$ vs $E_2$, with $\theta_x$~=~24$^\circ$ and $\theta_y$~=~45$^\circ$ corresponding to Fig.~\ref{ECR2D}. For each event total energy is obtained as 
\begin{equation}
E_{\rm tot}=E_1+E_2.
\end{equation}

\begin{figure}[H]
\centering
\includegraphics[scale=0.32]{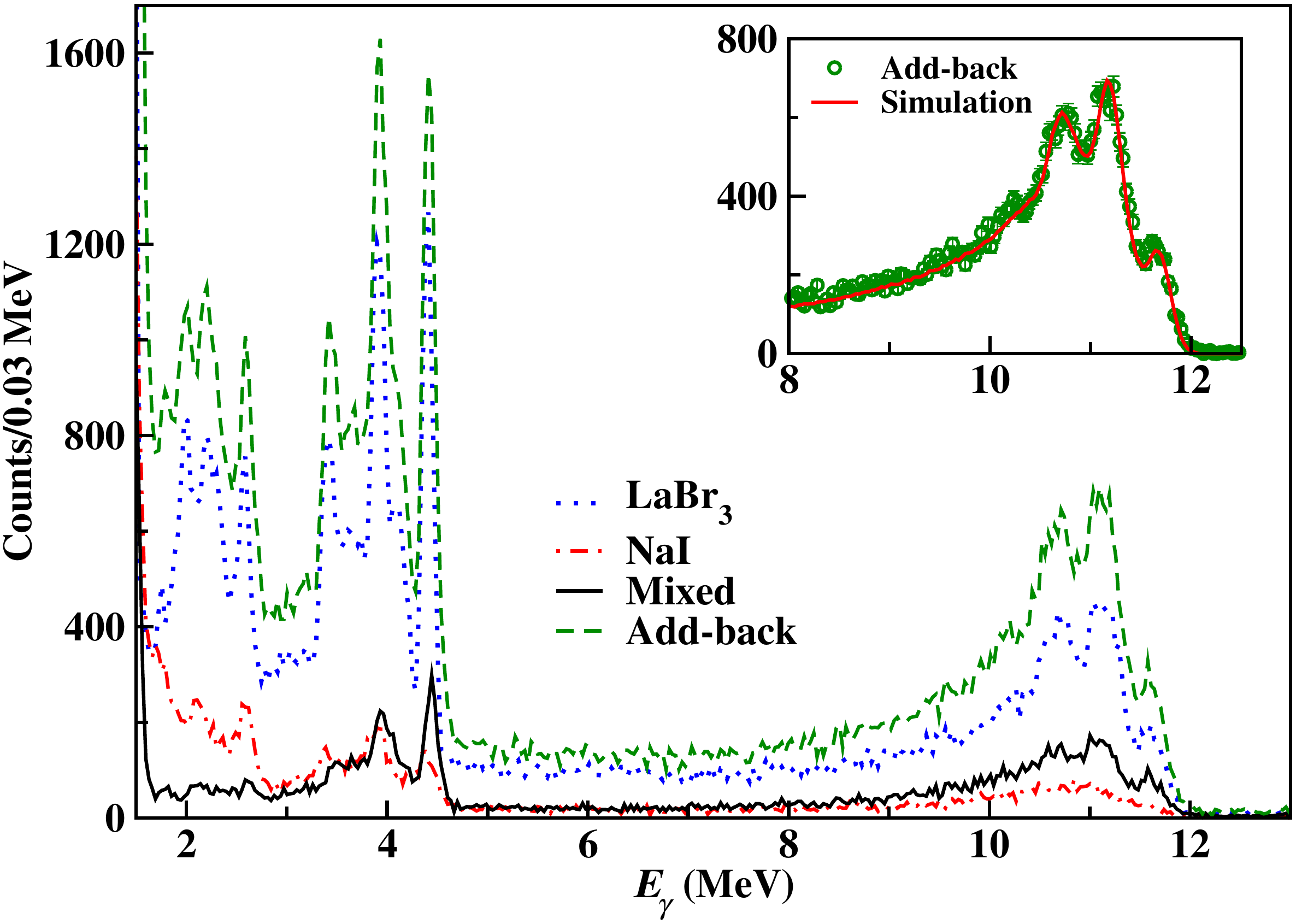}
\caption{\label{addback2}(Color online) The  energy spectra of D2 detector in $^{11}B(p,\gamma$) reaction at $E_p$~=~163~keV for pure LaBr$\rm _3(Ce)$ events (blue dotted line), pure NaI(Tl) events (red dash-dotted line) and  mixed events (black continuous line). The total energy spectra after add-back (green dashed line) is also shown for comparison. The inset shows a comparison of the experimental and the simulated add-back spectra for $E_{\gamma}$~=~11.7~MeV, where the spectra are normalized to total counts in the range $E_{\gamma}$~=~10.4 to 12.0 MeV.}
\end{figure}

Figure~\ref{addback2} shows the energy spectra of D2 detector from the $^{11}B(p,\gamma$) reaction at $E_p$ = 163~keV for pure LaBr$\rm _3(Ce)$ events, pure NaI(Tl) events and mixed events. The total energy spectra after add-back (green dashed line) is also shown for comparison, where a significant gain in integral counts due to add-back is evident both at 4.4 MeV and 11.7 MeV. 
 The enhancement factor (N($E_{\rm tot}$)/N$(E_{1}$)) is found to be 1.42 for the 4.4 MeV photo-peak and 1.52 for $E_{\gamma}$~=~10.4--12.0~MeV. 
It can be seen that the counts in NaI(Tl) are much smaller as compared to the $\rm LaBr_3(Ce)$ as expected. 
The simulated add-back spectra were generated following a procedure identical to that of the data. For each event, total energy is obtained by adding individual energy components after folding in respective resolution function for LaBr$_3$(Ce) and NaI(Tl) crystals.  A comparison of the experimental and the simulated add-back spectra for 11.7 MeV is shown in the inset of Fig.~\ref{addback2}, where the spectra are normalized to total counts in the range $E_{\gamma}$~=~10.4 to 12.0 MeV. The excellent agreement is clearly seen.
For high energy events, larger fraction of energy is in LaBr$_3$(Ce) and the add-back spectrum is therefore dominated by LaBr$_3$(Ce) energy component. In the add-back process the smaller energy component deposited in NaI(Tl) is recovered, resulting in increased counts in second-escape, first-escape and photo-peak as seen in both data and simulations, without any significant effect on the resolution.
 The energy resolution at 4.4~MeV after add-back is found to be $\sim$ 3.4\% as compared to $\sim$~2.9\% for  $\rm LaBr_3(Ce)$ crystal alone. 

\section{Summary and Conclusions}
Detailed characterization of the individual detector elements   ($\rm LaBr_3(Ce)$-NaI(Tl)) of PARIS detector has been carried out. 
The detector response is investigated over a wide range of $E_\gamma$ using radioactive sources and employing $^{11}B(p,\gamma$) reaction at $E_p$ = 163 keV and $E_p$ = 7.2 MeV  with CAEN digitizers.

The linearity of energy response of the $\rm LaBr_3(Ce)$ in the phoswich detector is tested upto $E_\gamma$~=~22.6~MeV using three different voltage dividers.  
 Stability of high voltage unit is of paramount importance as it can impact the count rate related drifts.
The detector gain stability is studied for count rates between 2--220~kHz using B3 type voltage divider (developed by IPHC Strasbourg group for the PARIS detector), for high voltages corresponding to linear energy reposnse upto 22.6 MeV. Detector performance was found to be stable over the count rate 50 to 150~kHz, where the centroid shift and broadening of FWHM at 835~keV is less than 1.5\% (11~keV) and 13\% (5~keV), respectively. The performance of B3 type voltage divider is found to be very good in terms of resolution, linearity of energy calibration and the count rate handling capability. The stability of the detector gain is found to be better than $\pm1\%$ over a period of 24 hours.

Time resolution of the phoswich detector is measured with $\rm ^{60}Co$ source after implementing CFD algorithm in the WAVEDUMP acquisition software. The FWHM (digital)~$\sim$~457~$\rm \pm$~3 ps is similar to FWHM (analog)~$\sim$~454~$\rm \pm$~6 ps, at the optimum PMT voltage ($E_{\rm full-scale}\sim$~2~MeV). The measured time resolution of individual detector FWHM~$\sim$~315~ps, is excellent for neutron-$\gamma$ separation by TOF measurement at close distances.
 At lower bias voltages corresponding to linear energy response upto high energy, the time resolution worsens by about 12\% (513~$\rm \pm$~3~ps). The energy resolution of $\sim$~2.1\% at 22.6~MeV is measured for configuration giving the best linearity upto high energy. The optimal configuration corresponding to A1~$\sim$~60~mV is found to give energy resolution better than 2\% at 22.6 MeV and non-linearity of $\sim$~10\% around 22~MeV. The detector efficiency is measured for $E_{\gamma}$~=~0.662~--~4.4~MeV for three different distances from the face of the detector. The measured efficiency with radioactive sources are in good agreement with GEANT4 based simulations. The relative intrinsic efficiency of $\rm LaBr_3(Ce)$ crystal at 11.7~MeV with respect to that at 4.4~MeV is found to be $\sim$~24\%. The lineshape of high energy gamma rays has been reproduced by GEANT4 based simulation. It is demonstrated that the $ E_{\rm tot}$ spectrum with the add-back of individual components of the phoswich detector gives substantial efficiency gain of about 50\%, without significantly affecting the resolution. Further, the present measurements indicate that for the composite signal of the phoswich detector, the digital signal processing  with 0.5--1~GHz sampling rate and high resolution (12--14 bit) ADC will be optimal.

\acknowledgments
We thank Mr.~K.V.~Divekar for assistance during the measurements,  Prof.~L.C.~Tribedi and Mr. Shamik Bhattacharjee for help in the ECR based experiment  and Dr. Namita Maiti for the Boron target. We are thankful to the ECR staff and the PLF staff for smooth machine operation during proton beam experiments. The support from PARIS collaboration is gratefully acknowledged.

\end{document}